\documentclass[aps,prl,twocolumn,superscriptaddress]{revtex4-2}

\usepackage[dvipdfmx]{graphicx}
\usepackage{amsmath}
\usepackage{amssymb}
\usepackage{bm}
\usepackage{url}

\newcommand{\degree}{^\circ}

\begin{document}

%\preprint{}

\title{Observational Evidence for Stochastic Shock Drift Acceleration of Electrons at the Earth's Bow Shock}
%\title{}


\author{T. Amano}
\affiliation{Department of Earth and Planetary Science, University of Tokyo, Tokyo, Japan.}
\email[]{amano@eps.s.u-tokyo.ac.jp}

\author{T. Katou}
\affiliation{Department of Earth and Planetary Science, University of Tokyo, Tokyo, Japan.}

\author{N. Kitamura}
\affiliation{Department of Earth and Planetary Science, University of Tokyo, Tokyo, Japan.}

\author{M. Oka}
\affiliation{Space Sciences Laboratory, University of California, Berkeley, CA, USA}

\author{Y. Matsumoto}
\affiliation{Department of Physics, Chiba University, Chiba, Japan.}

\author{M. Hoshino}
\affiliation{Department of Earth and Planetary Science, University of Tokyo, Tokyo, Japan.}

\author{Y. Saito}
\affiliation{Institute of Space and Astronautical Science, Sagamihara, Japan.}

\author{S. Yokota}
\affiliation{Department of Earth and Space Science, Osaka University, Toyonaka, Japan.}

\author{B. L. Giles}
\affiliation{NASA Goddard Space Flight Center, Greenbelt, MD, USA.}

\author{W. R. Paterson}
\affiliation{NASA Goddard Space Flight Center, Greenbelt, MD, USA.}

\author{C. T. Russell}
\affiliation{Department of Earth, Planetary, and Space Sciences, University of California, Los Angeles, CA, USA.}

\author{O. Le Contel}
\affiliation{Laboratoire de Physique des Plasmas, CNRS/Ecole Polytechnique/Sorbonne Université/Univ. Paris-Sud/Obs. de Paris, Paris, France.}

\author{R. E. Ergun}
\affiliation{Laboratory for Atmospheric and Space Physics, University of Colorado, Boulder, CO, USA.}

\author{P.-A. Lindqvist}
\affiliation{Royal Institute of Technology, Stockholm, Sweden.}

\author{D. L. Turner}
\affiliation{Space Sciences Department, The Aerospace Corporation, CA, USA.}

\author{J. F. Fennell}
\affiliation{Space Sciences Department, The Aerospace Corporation, CA, USA.}

\author{J. B. Blake}
\affiliation{Space Sciences Department, The Aerospace Corporation, CA, USA.}

%\thanks{}


\date{\today}

\begin{abstract}
The first-order Fermi acceleration of electrons requires an injection of electrons into a mildly relativistic energy range. However, the mechanism of injection has remained a puzzle both in theory and observation. We present direct evidence for a novel stochastic shock drift acceleration theory for the injection obtained with Magnetospheric Multiscale (MMS) observations at Earth's bow shock. The theoretical model can explain electron acceleration to mildly relativistic energies at high-speed astrophysical shocks, which may provide a solution to the long-standing issue of electron injection.
\end{abstract}

%\keywords{}

\maketitle

Radio and X-ray synchrotron emissions from young supernova remnant (SNR) shocks indicate that they are efficient acceleration sites of cosmic-ray electrons \cite{Koyama1995,Bamba2003a}. By contrast, the observations of non-thermal electrons associated with shocks are rare in the heliosphere \cite{Dresing2016}. Conversely, proton acceleration appears to be common both in heliospheric and SNR shocks, suggesting more efficient acceleration than electrons. It is believed that the diffusive shock acceleration (DSA) process \cite{Blandford1987} is responsible for the acceleration of both species, but perhaps with different efficiencies of injection into the process.

It is known that mildly relativistic energies are typically needed for the electron injection, which is much more stringent than protons, indicating that the injection process may be activated only at stronger SNR shocks. The apparent discrepancy in the observed electron acceleration efficiency between heliospheric and SNR shocks may at least partially be attributed to the suspected dependence of electron injection efficiency on parameters of the shock. Despite extensive theoretical \cite{Levinson1992,Amano2010}, numerical \cite{McClements2001,Hoshino2002,Amano2007,Riquelme2011,Matsumoto2015,Matsumoto2017}, and observational \cite{Gosling1989,ShimadaN.1999,Oka2006,OkaM.2009,Masters2013,Wilson2016,Masters2016} studies over the decades, this problem of electron injection has remained unresolved. We here present in-situ measurements of Earth's bow shock by NASA's four-spacecraft Magnetospheric Multiscale (MMS) mission \cite{J.L.Burch2016}, which provides the direct evidence for a recently-proposed model called stochastic shock drift acceleration (SSDA) \cite{Katou2019a} as the mechanism of sub-relativistic electron acceleration.

\begin{figure}
 \includegraphics[width=8.6cm]{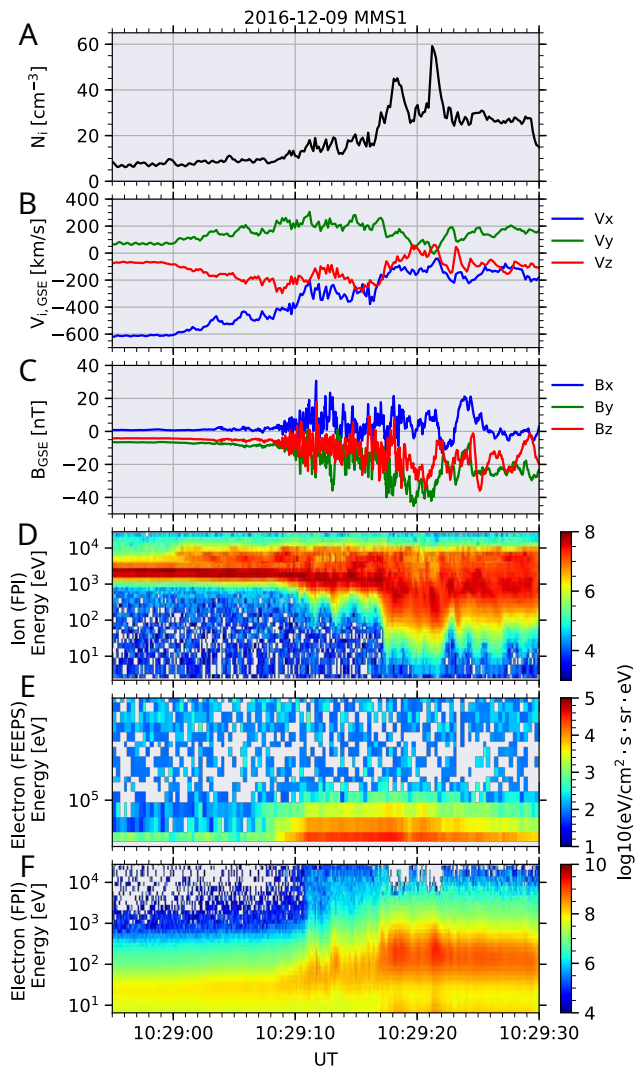}%
 \caption{Overview of Earth's bow shock crossing on 2016 December 9 observed by the MMS1. (A) Ion density. (B) Ion bulk velocity. (C) Magnetic field. (D-F) Energy-time spectrogram in units of differential energy flux for (D) ions, (E) high energy electrons measured by FEEPS, and (F) low energy electrons measured by FPI, respectively. The vector quantities are expressed in the GSE coordinate.
\label{fig1}}
\end{figure}

On 2016 December 9 around 10:29 UT, MMS crossed Earth's bow shock from the upstream (solar wind) to the downstream (magnetosheath). Fig.~\ref{fig1} shows an overview of the observation by MMS1. A gradual density compression and deceleration of anti-sunward plasma flow speed started around 10:29:00 UT, indicating that the spacecraft entered into the shock transition layer (STL). At around 10:29:09 UT, MMS started to measure substantial magnetic field fluctuations. Energetic electron ($\gtrsim$ 1 keV) flux enhancements began nearly simultaneously. The flow deceleration continued up to around 10:29:20 UT when the magnetic field magnitude exhibited a peak. We interpret this as an encounter of the magnetic overshoot, and the region following is the downstream.

We estimated a shock normal vector $\bm{n} = (+0.979, -0.827, -0.525)$ in the Geocentric Solar Ecliptic (GSE) coordinate using a method similar to \cite{Vinas1986}. We then obtained the solar wind speed normal to the shock surface in the spacecraft frame $u_0 \simeq 590$ km/s, and the corresponding upstream Alfven Mach number $M_{\rm A} \simeq 8.9$. Similarly, the magnetic field obliquity (angle between the upstream magnetic field and the shock normal) was estimated to be $\theta_{Bn} \simeq 85\degree$. In addition, we estimated the shock propagation speed with respect to the spacecraft as $U_0 \simeq 30 $ km/s using the method based on the thickness of the shock foot \cite{Livesey1984,Gosling1985}.

\begin{figure}
 \includegraphics[width=8.6cm]{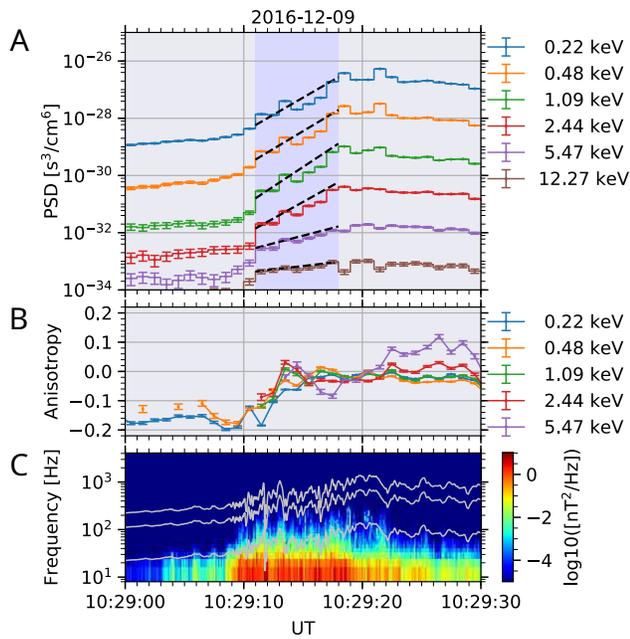}%
 \caption{Evidence for electron diffusion in shock transition layer. (A) Phase space density (PSD) of electrons. (B) First-order pitch-angle anisotropy of electrons. (C) Frequency-time spectrogram of magnetic field fluctuations. Three solid lines in (C) indicate, from top to bottom, $f_{\rm ce}$, $f_{\rm ce}/2$, $f_{\rm ce}/10$ where $f_{\rm ce}$ denotes the electron cyclotron frequency. The black dashed lines in (A) represent fitting results with an exponential function. The shaded time interval was used for the fitting.
 \label{fig2}}
\end{figure}

The energetic electron intensities shown in Fig.~\ref{fig2}A for selected energy channels measured by the FPI (Fast Plasma Investigation) instruments exhibited exponential increases toward downstream between 10:29:10 and 10:29:20 UT. (Note that we used all four spacecraft data and took average over one second to increase the statistics.) If the particles perform diffusion in space that is balanced with the convection, such an exponential profile will result. The diffusion approximation is valid only when efficient isotropization of pitch-angle distribution is realized. It is usually a reasonable description for high energy particle transport at large spatial scales. However, the observation implies that it applies for low-energy electrons on the scale size of the thin shock layer.

Under the assumption of weak anisotropy, we can indeed derive a diffusion-convection equation from the so-called focused transport equation for the gyrotropic distribution function $f(v,\mu)$ (where $v$ and $\mu$ denote the particle speed and pitch angle cosine) \cite{Skilling1975,Zank2014b}. Using the Legendre polynomial expansion for the pitch-angle distribution $f(v,\mu) = \sum_{n=0}^{\infty} (n+1/2) g_n(v) P_n(\mu)$ and assuming $|g_0| \gg |g_1|, |g_2|$ (weak anisotropy), we obtain the following diffusion-convection equation for the isotropic part $g_0$ of the distribution function
\begin{align}
\frac{\partial g_0}{\partial t} +
u_{\rm sh} \frac{\partial g_0}{\partial s} +
\frac{1}{3} \frac{\partial \ln B}{\partial s} \frac{\partial g_0}{\partial \ln v} =
\frac{\partial}{\partial s} \left( \kappa \bm{b} \frac{\partial g_0}{\partial s} \right),
\label{eq:difconv}
\end{align}
from Eq.(6) of \cite{Katou2019a} that describes the electron transport within a quasi-perpendicular STL ($\cos \theta_{Bn} \ll 1$) in the de Hoffmann-Teller frame (HTF). Here, $s, \kappa, \bm{b}$ are respectively the spatial coordinate along the magnetic field, parallel diffusion coefficient, and the magnetic field unit vector. Note that $u_{\rm sh} = u_0/\cos \theta_{Bn}$ gives the upstream flow speed in the HTF. The diffusion coefficient $\kappa$ is related to the pitch-angle scattering rate $D_{\mu\mu}$ via $\kappa = v^2/(6 D_{\mu\mu})$, which is obtained by assuming $D_{\mu\mu}$ is independent of $\mu$. Alternatively, it may be understood as the scattering rate averaged over pitch angle, which eliminates any pitch-angle dependence as long as the anisotropy is sufficiently small.

The first-order pitch-angle anisotropy defined as $g_1/g_0$ is shown in Fig.~\ref{fig2}B (see, Supplemental Material, for detail). The anisotropies for supra-thermal ($\sim 0.2{\text -}1$ keV) electrons were consistently negative before 10:29:10 UT, meaning that these electrons were nearly free streaming anti-parallel to the magnetic field. We think that they were escaping from the shock toward upstream, which is consistent with the upstream magnetic field that was directed toward downstream. The anisotropies started to decline at around 10:29:10 UT to become nearly isotropic, which then continued in the remainder of the transition region (up to around 10:29:20 UT). We also confirmed $|g_2/g_0| \ll 1$ during this time interval (see, Supplemental Material), which indicates that the electron transport is adequately described by Eq.~(\ref{eq:difconv}).

The observed nearly isotropic distribution should be kept only when strong pitch-angle scattering by plasma waves operates. Indeed, electromagnetic fluctuations were substantially enhanced during the same time interval (Fig.~\ref{fig2}C). In particular, we confirmed the appearance of intense high-frequency and coherent whistler waves with right-hand polarization \cite{Zhang1999,Hull2012,Oka2017} which can scatter low-energy electrons via cyclotron resonance (see, Supplemental Material, for polarization analysis result \cite{Santolik2003}).

\begin{figure}
 \includegraphics[width=8.6cm]{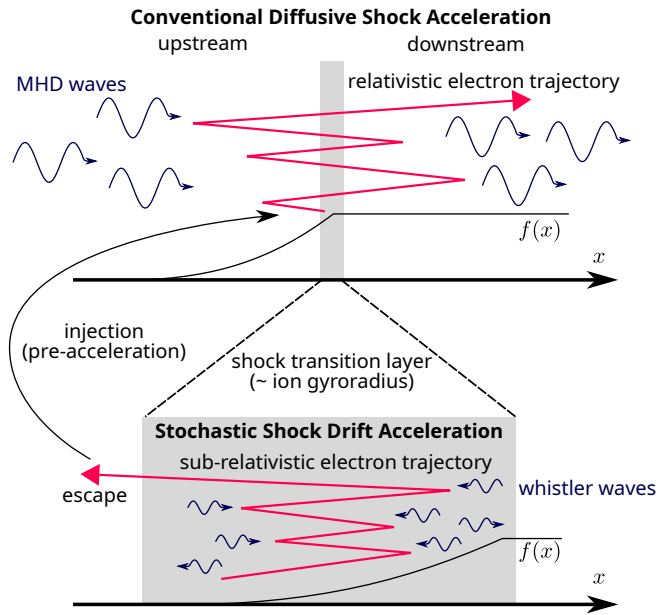}%
 \caption{Schematics illustrating relation between conventional diffusive shock acceleration (DSA) and stochastic shock drift acceleration (SSDA) models. Particle acceleration via DSA operates in a spatial extent much larger than the thickness of the shock. In contrast, electrons accelerated by SSDA are confined within the shock transition layer with a typical thickness of ion gyroradius. DSA usually assumes that particles are scattered by MHD waves, which is possible for electrons only if they have relativistic energies. SSDA may accelerate sub-relativistic electrons because of intense whistler waves in the transition layer, and may potentially provide a seed population for DSA.
 \label{fig3}}
\end{figure}

As we have seen, the simultaneous appearance of the all three features (the exponential particle intensity, near isotropy, and enhanced fluctuations) validates the diffusion approximation for the electron transport in the STL. However, this does not necessarily mean that the energetic electrons were accelerated by the standard DSA. This is because the particle acceleration was apparently occurring within the STL with a finite thickness, which is neglected in the standard theory. In contrast, the observation agrees quite well with the picture provided by the theory of SSDA \cite{Katou2019a}, which is schematically illustrated in Fig.~\ref{fig3}.

In the conventional shock drift acceleration (SDA) model \cite{Wu1984,LeroyM.M.1984}, electrons staying in the STL experience a nearly constant rate of energy gain by performing the magnetic-field gradient drift in the direction anti-parallel to the convection electric field ($E = -V \times B$). A finite interaction time between the particles and the shock limits the energy gain by SDA. Recent three-dimensional ab initio modeling of collisionless shocks \cite{Matsumoto2017} as well as in-situ measurements by MMS \cite{Oka2017} both found that pitch-angle scattering, which is not taken into account in SDA, can be efficient in the STL. The SSDA theory thus introduces stochastic pitch-angle scattering of electrons, which diffusively confine the particles within the acceleration region longer than in the absence of scattering. It is important to point out that the scattering is essential for the particle confinement, but the energization mechanism itself remains the same as the classical SDA, i.e., due to the DC electric field acceleration. Indeed, as we see in Eq.~(\ref{eq:difconv}), the particle energy gain is simply proportional to the magnetic field gradient and independent of wave properties. We note that the energy gain associated with the flow velocity divergence (the major energy gain for the standard DSA) is negligible at quasi-perpendicular shocks \cite{Katou2019a}.

In contrast to the SDA, the efficient confinement leads to the formation of a power-law spectrum with a cutoff at high energy that is determined by the rate of pitch-angle scattering $D_{\mu\mu}$. More specifically, the cutoff in the power-law appears when the particle diffusion length becomes longer than the size of the acceleration region, because in this case the particles can no longer be confined in the system. We thus expect that the power-law will form only when $D_{\mu\mu}$ is larger than the theoretical threshold \cite{Katou2019a}
$$
\frac{D_{\mu\mu}^{*}}{\Omega_{\rm ce}} =
\frac{1}{6 \eta}
\Bigl( \frac{m_{\rm e}}{m_{\rm i}} \Bigr)
\Bigl( \frac{E}{E_{\rm sh}} \Bigr),
$$
where $\Omega_{\rm ce}$, $E$ are the electron cyclotron frequency, and the electron kinetic energy. $E_{\rm sh} = (1/2) m_{\rm e} u_{\rm sh}^2$ represents the upstream electron flow kinetic energy in the HTF. We have introduced a constant numerical factor $\eta$ which is defined such that $\eta u_0/\Omega_{\rm ci}$ (rather than the entire shock thickness estimated by the ion gyroradius $\sim u_0/\Omega_{\rm ci}$) gives the spatial extent of the electron acceleration region. In other words, it represents a fractional thickness in which the wave activity is high allowing the efficient confinement of the particles. By assuming a quasi-steady shock structure passing through the spacecraft with a constant speed, we estimated $\eta \sim 0.4$ as the ratio between the time intervals of the acceleration region ($\sim 10$ s) and the entire STL ($\sim 25$ s). Note that the factor $1/(6\eta)$ was missing in the order of magnitude estimate by \cite{Katou2019a}. Although the power-law index found in the observation was roughly consistent with the theoretical prediction \cite{Katou2019a}, the power-law form is quite common in space plasma environments and this fact itself may not necessarily provide convincing evidence for the theory. On the other hand, given the macroscopic shock parameters, the scattering rate is the only parameter that controls the cutoff energy. Therefore, this property can be used to prove the theory.

\begin{figure}
 \includegraphics[width=8.6cm]{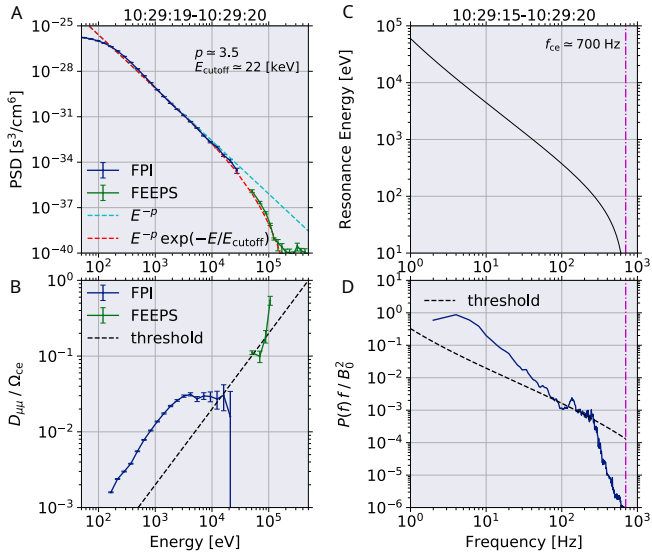}%
 \caption{Comparison between theory and observation. (A) Electron phase space density (PSD) averaged over one second (from 10:29:19 to 10:29:20 UT). The red dashed line represents a fitting result with a model function. (B) Scattering rate estimated from the particle intensity profiles. (C) Minimum resonance energy. (D) Magnetic field power spectrum averaged over five seconds (from 10:19:15 to 10:29:20 UT). The magenta dash dotted lines in (C), (D) represent the electron cyclotron frequency. The theoretical threshold shown with the black dashed lines in (B) and (D) were below the measurements in the energy range where the power-law was observed.
 \label{fig4}}
\end{figure}

The scattering rate as a function of energy $D_{\mu\mu} (E)$ may be estimated by fitting the particle intensity profiles using the function $f(E,t) \propto \exp \left[ t/\tau(E) \right]$ where the fitting parameter $\tau(E)$ is the e-folding time as a function of energy. Examples of the fitting results are shown in Fig.~\ref{fig2}A. Assuming again that the shock structure was stationary during the shock crossing, we converted $\tau(E)$ to $D_{\mu\mu} (E)$ using the formula
$$
\frac{D_{\mu\mu}}{\Omega_{\rm ce}} =
\frac{1}{6}
\Bigl(
\frac{m_{\rm e}}{m_{\rm i}}
\Bigr)
\Bigl(
\frac{E}{E_{\rm sh}}
\Bigr)
\Bigl(
\frac{u_0/\Omega_{\rm ci}}{U_0 \tau(E)}
\Bigr)
.
$$
This may be obtained by equating the observed spatial scale $U_0 \tau(E)$ and the diffusion length normal to the shock $\kappa \cos \theta_{Bn} / u_{\rm sh}$ as implied by Eq.~(\ref{eq:difconv}).

The electron energy spectrum and the estimated $D_{\mu\mu} (E)$ are shown in Fig.~\ref{fig4}A and \ref{fig4}B, respectively. The results obtained for higher-energy electrons measured by the FEEPS (Fly's Eye Energetic Particle Spectrometer) instruments are also shown. Comparison between the estimated scattering rate and the theoretical threshold (the black dashed line) indicates that the cutoff should appear at around $\sim 20$ keV. We determined an exponential cutoff energy of $E_{\rm cutoff} \simeq 22 \pm 1$ keV by fitting the energy spectrum with the function $f(E) \propto E^{-p} \exp(-E/E_{\rm cutoff})$ \cite{Oka2006}. Considering that the theoretical cutoff is an order of magnitude estimate, we see that the agreement between the theory and observation is quite good.

Another independent test is to use a quasi-linear theory that relates the fluctuation power spectrum to the scattering rate. We then obtain the threshold wave power
$$
\frac{P^{*}(f) f}{B_0^2} =
\frac{2}{3 \pi \eta}
\Bigl(
\frac{m_e}{m_i}
\Bigr)
\Bigl(
\frac{B}{B_0}
\Bigr)
\Bigl(
\frac{E}{E_{\rm sh}}
\Bigr)
\left[
\Bigl(1 + \left| \frac{\omega}{k v \mu} \right| \Bigr)
\frac{d \ln \omega}{d \ln k}
\right]^{-1},
$$
which includes corrections into \cite{Katou2019a} due to a finite wave frequency in the resonance condition and the dispersive effect. To be consistent with the SSDA theory, the measured power must be larger than the threshold in the frequency range where the power-law energy spectrum was formed in the corresponding resonance energy range. Fig.~\ref{fig4}C shows the relation between the wave frequency and the minimum energy of the particles that can be scattered by the wave via the cyclotron resonance. (We used the cold plasma dispersion relation for right-hand circularly polarized parallel propagating waves for the calculation of resonance energy.) Fig.~\ref{fig4}D compares the measured power spectrum and the theoretical threshold. Note that the actual threshold may be less stringent because the coherent nature of high-frequency whistlers potentially leads to more efficient scattering than this estimate. We thus conclude that the measured wave power is at least the same level and perhaps larger than that required to account for the electron scattering.

In general, the most stringent condition for the wave power will be imposed at high frequency ($\gtrsim 0.1 f_{\rm ce}$ where $f_{\rm ce} = \Omega_{\rm ce}/(2\pi) \simeq 700$ Hz) because the power falls off rapidly as the frequency increases. The whistler waves in this frequency range can scatter electrons with energies $\sim 0.1{\text -}1$ keV via the cyclotron resonance. As increasing the energy, the electrons start to interact with lower-frequency larger-amplitude waves. They are often in oblique propagation and can scatter the particles much more efficiently through Landau, transit time, or higher harmonic cyclotron resonances \cite{Oka2019}, although the efficiency will saturate at some point due to nonlinearity $\delta B \sim B$ as was seen at a few keV in Fig.~\ref{fig4}B. This strongly indicates that the intensity of high-frequency whistlers is the crucial ingredient for the acceleration of non-thermal electrons. Unless they have sufficiently large power, the production of non-thermal electrons will not be triggered in the first place. Note that the theoretical threshold wave power has a strong dependence on the Alfven Mach number and magnetic-field obliquity $\propto (M_{\rm A} / \cos \theta_{Bn})^2$. Therefore, even with the same level of wave power, the non-thermal production rate may substantially change depending on the shock parameters.

The injection has been the central issue in the shock acceleration theory. The lack of an efficient mechanism for generating high-frequency whistler waves makes the electron injection much more difficult than the ion injection. What has not been taken into account so far is that the indispensable scattering agent exists only within a thin layer. The generation of high-frequency whistler waves is probably related to the pitch-angle anisotropy. We confirmed that weak but clear perpendicular anisotropy ($g_2/g_0 < 1$) developed at around $\sim 1$ keV in the electron acceleration region. Such anisotropy, naturally produced by the adiabatic SDA (heating due to magnetic field compression), may destabilize high-frequency whistler waves via the electron cyclotron resonance \cite{Amano2010,Tokar1984}. The generated waves induce scattering that, if it is strong enough, will transform SDA into SSDA, suggesting that the particle acceleration by SSDA may proceed in a self-sustaining manner. We note that the anisotropy at the highest time resolution (30 ms) showed substantial time variability, suggesting that the competition between the production and relaxation of anisotropy was occurring in a highly dynamic manner. Once the process is triggered in the lowest energy where the power-law starts to form, higher energy particles are scattered more easily by more intense lower-frequency fluctuations.

In summary, we have demonstrated that the two completely independent measurements (Fig.~\ref{fig4}B and \ref{fig4}D) both gave quantitatively consistent results with the theoretical prediction, which thus provides strong support for the SSDA model. This finding has important implications for astrophysical shocks. The theoretical scaling law \cite{Katou2019a} combined with the estimated scattering rate suggests that the cutoff energy is given by
$$
E_{\rm cutoff} \simeq 20 \, {\rm keV} \,
\Bigl( \frac{u_0}{600 \, {\rm km/s}} \Bigr)^2
\Bigl( \frac{\cos \theta_{Bn}}{\cos 85 \degree} \Bigr)^{-2}
\Bigl( \frac{D_{\mu\mu}}{0.03 \Omega_{\rm ce}} \Bigr).
$$
Therefore, shocks in the heliosphere (with a typical shock speed of $\sim 400$ km/s) will not normally produce relativistic electrons, and subsequent DSA will not take place. In contrast, high-speed ($\gtrsim$ 3000 km/s) young SNR shocks with a relatively large obliquity will accelerate electrons to more than a few hundreds of keV within the STL. The pre-accelerated electrons will be further energized to ultra-relativistic energies by DSA that operates in a much larger spatial extent. The strong dependence of the cutoff energy on the obliquity implies that the turbulence in the upstream of the shock is an important factor controlling the number of injected electrons. For instance, large-amplitude magnetic field fluctuations ahead of a quasi-parallel shock can locally produce a portion of the shock which behaves as a nearly perpendicular shock \cite{Balogh2013}. Therefore, the macroscopic injection efficiency in the actual astrophysical environment will be determined as a result of the nonlinear dynamical evolution of collisionless shocks. This point may be important to understand the relation between the quasi-perpendicular injection scenario proposed here and apparent radial magnetic fields as inferred from polarization measurements of radio synchrotron emission from young SNRs \cite{Reynoso2013,West2017}.

\begin{acknowledgments}
This work was supported by JSPS KAKENHI Grants No. 17H02966 and 17H06140. We appreciate J.~L.~Burch and the entire MMS team for the design and operation of the mission. The data used in this study were obtained by FPI (Fast Plasma Investigation) \cite{Pollock2016}, FEEPS (Fly's Eye Energetic Particl Spectrometer) \cite{Blake2016}, FGM (Flux Gate Magnetometer) \cite{Russell2016}, SCM (Search Coil Magnetometer) \cite{LeContel2016}, and EDP (Electric-field Double Probe) \cite{Ergun2016,Lindqvist2016}. All the MMS data are publicly available at the MMS Science Data Center \url{https://lasp.colorado.edu/mms/sdc/public/}.
\end{acknowledgments}

%\bibliography{library}

\end{document}